\documentclass[conference,letterpaper]{IEEEtran}
\usepackage[letterpaper, left=0.65in, right=0.65in, bottom=1.1in, top=0.75in]{geometry}
\IEEEoverridecommandlockouts
\usepackage{amsthm,amsmath,amssymb,amsfonts,mathtools}
\usepackage{tabu}
\usepackage{graphicx}
\usepackage{xcolor,soul}
\usepackage{subcaption}
\usepackage{longtable}
\usepackage{booktabs}
\usepackage{lipsum}
\usepackage{algorithm}
\usepackage{algpseudocode}
\usepackage{setspace}
\usepackage{multirow}
\usepackage{svg}
\usepackage{cite}

\DeclareMathOperator{\diag}{diag}

\begin{document}

\title{Channel Estimation for Reconfigurable Intelligent Surface MIMO with Tensor Signal Modelling}

% \author{\IEEEauthorblockN{Alexander James Fernandes and Ioannis Psaromiligkos} \\
% \thanks{{Department of Electrical and Computer Engineering}, 
% {McGill University}, Montreal, QC, Canada. Email: 
% alexander.fernandes@mail.mcgill.ca;\\ ioannis.psaromiligkos@mcgill.ca}
% \thanks{This work was supported in part by the Natural Science and Engineering Research Council of Canada under the Discovery Grant Program and in part by the Vadasz Scholar McGill Engineering Doctoral Award.}
% }

\author{\IEEEauthorblockN{Alexander James Fernandes and Ioannis Psaromiligkos}
\IEEEauthorblockA{Department of Electrical and Computer Engineering, McGill University, Montreal, QC, Canada}
}

% The paper headers
\markboth{}%
%\markboth{Name of Journal,~Vol.~\#, No.~\#, Month~Year}%
{}

\maketitle

\begin{abstract}
We consider a narrowband MIMO reconfigurable intelligent surface (RIS)-assisted wireless communication system and use tensor signal modelling techniques to individually estimate all communication channels including the non-RIS channels (direct path) and decoupled RIS channels.
We model the received signal as a third-order tensor composed of two CANDECOMP/PARAFAC decomposition terms for the non-RIS and the RIS-assisted links, respectively, and we propose two channel estimation methods based on an iterative alternating least squares (ALS) algorithm:
The two-stage RIS OFF-ON method estimates each of the non-RIS and RIS-assisted terms in two pilot training stages, whereas the enhanced alternating least squares (E-ALS) method improves upon the ALS algorithm to jointly estimate all channels over the full training duration.
A key benefit of both methods compared to the traditional least squares (LS) solution is that they exploit the structure of the tensor model to obtain decoupled estimates of all communication channels.
We provide the computational complexities to obtain each of the channel estimates for our two proposed methods.
Numerical simulations are used to evaluate the accuracy and verify the computational complexities of the proposed two-stage RIS OFF-ON, and E-ALS, and compare them to the traditional LS methods.
Results show that E-ALS will obtain the most accurate estimate while only having a slightly higher run-time than the two-stage method.
\end{abstract}


\section{Introduction}
In recent years, the reconfigurable intelligent surface (RIS), has been gaining interest as an effort to improve the capabilities of the future of wireless communication systems \cite{Hassouna2023}.
The RIS is a 2D surface composed of several passive reflective elements with the ability to redirect the path of reflection electronically which allows to spatially focus a line-of-sight (LOS) path between the transmitter and receiver locations.
One major challenge with RIS-assisted communication systems is estimating the channel state information (CSI) due to having a large number of channels going through the RIS.

A powerful tool that has advanced the field of multi-sensor signal processing in the past decade are multiway arrays involving tensor signal modelling and decompositions \cite{Kolda2009}.
In particular CANDECOMP/PARAFAC (CP) decomposition is a popular method for channel estimation in RIS systems due to being able to estimate the decoupled channels going through the RIS \cite{DeAraujo2023,Wei2021,DeAraujo2021,DeAraujo2020,Xiao2022,Wei2021a,Wei2020,Yuan2022,Yang2022a,Beldi2023,Yuan2022a,Li2023,Hashi2022,Sokal2023}.
These studies focus on estimating the RIS-assisted channels and most assume the non-RIS-assisted channel (i.e., the direct-path between the AP and UEs) is blocked.
Only the study by DeArujo \textit{et al.} considers the direct-path for channel estimation \cite{DeAraujo2023}.
In their study they consider semi-blind channel and symbol estimation using two transmission stages: 
in the first stage, the RIS is turned OFF to estimate the direct-path. 
In the second stage, the RIS is turned ON to estimate the symbols and RIS channels after subtracting the direct-path link estimate from the receive signal.
Due to estimation occurring in two transmission stages, we note that the pilots are not used efficiently to jointly estimate the direct-path and RIS channels such as what the least squares solution provides \cite{Jensen2020,Swindlehurst2022}.

In this paper we propose two channel estimation methods for RIS MIMO systems using tensor signal modelling to estimate all communication channels individually involving the direct-path, and RIS channels.
Our main contributions are:
\begin{itemize}
    \item We derive a new third-order tensor signal model for a narrowband RIS MIMO system composed of two CP decompositions: the non-RIS and the RIS-assisted path.
    \item We derive a two-stage channel estimation method that is similar to the semi-blind estimation method of \cite{DeAraujo2023} but without performing symbol estimation.
    In the first stage, we turn the RIS OFF to estimate the non-RIS assisted channels.
    In the second stage, we estimate the RIS-assisted path with an iterative alternating least squares (ALS) algorithm by turning the RIS ON and removing the non-RIS path from the receive signal using estimates obtained from the first stage.
    \item We propose another channel estimation method by enhancing the ALS algorithm to estimate both direct path and RIS channels jointly by deriving another iterative algorithm that exploits the structure of the tensor signal model composed of two CP decompositions.
    \item We provide computational complexities based on system model parameters for both channel estimation methods.
\end{itemize}
Finally, simulations evaluate the accuracy and the run-time of each method.
Overall, the E-ALS method is more accurate with higher complexity when jointly estimating the non-RIS and RIS-assisted paths.

The rest of the paper is organized as follows.
Section \ref{sec:tensorsigmodel} outlines the tensor RIS-assisted MIMO communication model and the block transmission scheme.
In Section \ref{sec:channel_estimation} we propose channel estimation methods based on an iterative ALS algorithm \cite{Kolda2009} and provide the computational complexities.
The accuracy and run-times are evaluated through simulations in Section \ref{sec:simulations}, and we conclude this paper in Section \ref{sec:conclusion}.

\textit{Notation:}
Column vectors are denoted as boldface lowercase ($\textbf{a}$), matrices as boldface uppercase ($\textbf{A}$), and scalars as uppercase ($A$) or lowercase ($a$).
Tensors are symbolized by calligraphic letters ($\mathcal{A}$).
The notation $[\textbf{A}]_{i,j}$ represents the element of the $i$-th row and $j$-th column of the matrix $\textbf{A}$.
A colon ($:$) is used to denote a placeholder for all elements in the given dimension, e.g., $[\textbf{A}]_{:,j}$ represents a column vector with all row elements from the $j$-th column of $\textbf{A}$.
The identity matrix of dimensions $N \times N$ is $\textbf{I}_N$, $\textbf{0}_{M \times N}$ is an $M \times N$ matrix of zeros, and $\textbf{1}_{M \times N}$ an $M \times N$ matrix of ones. 
Matrix operations on a matrix $\textbf{A}$ are denoted as: transpose $\textbf{A}^T$, conjugate transpose $\textbf{A}^H$, inverse $\textbf{A}^{-1}$, right pseudoinverse $\textbf{A}^\dagger_{\text{right}} = \textbf{A}^H (\textbf{A}\textbf{A}^H)^{-1}$ and left pseudoinverse $\textbf{A}^\dagger_{\text{left}} = (\textbf{A}^H \textbf{A})^{-1} \textbf{A}^H$.
A diagonal square matrix with the elements of a vector $\textbf{d}$ on its diagonal is expressed as $\diag(\textbf{d})$. 
The operator $\textbf{D}_{i}(\textbf{A})$ forms a diagonal matrix with the diagonal being the $i$-th row vector of $\textbf{A}$. 
The function $\text{vec}(\textbf{A})$ creates a vector by stacking the columns of $\textbf{A}$. 
The Frobenius norm of a matrix is $\|\textbf{A}\|_F$.
The matrix products are denoted as: Kronecker $\otimes$ and Khatri-Rao $\diamond$. 
A circular complex multivariate Gaussian distribution with mean $\boldsymbol{\mu}$ and covariance $\boldsymbol{\Sigma}$ is denoted as $\mathcal{CN}(\boldsymbol{\mu}, \boldsymbol{\Sigma})$.
Finally, the expectation operator is denoted as $\mathbb{E}[\cdot]$. 

\section{Tensor Signal Model}
\label{sec:tensorsigmodel}
The goal of this study is to estimate the CSI at the AP by transmitting pilots from the UE.
In this section, we first describe the system model, and then we describe the block transmission scheme, where a sequence of pilots from the UEs are retransmitted every block with a fixed duration while the RIS phase shifts are updated over every block.
Following this block transmission scheme, we remodel the received signal into a tensor composed of two CP decomposition terms.

\subsection{System Model}
\begin{figure}[t]
    \centering
    % \includesvg[width=\columnwidth]{SystemModel.svg}
    \includegraphics[width=0.9\columnwidth]{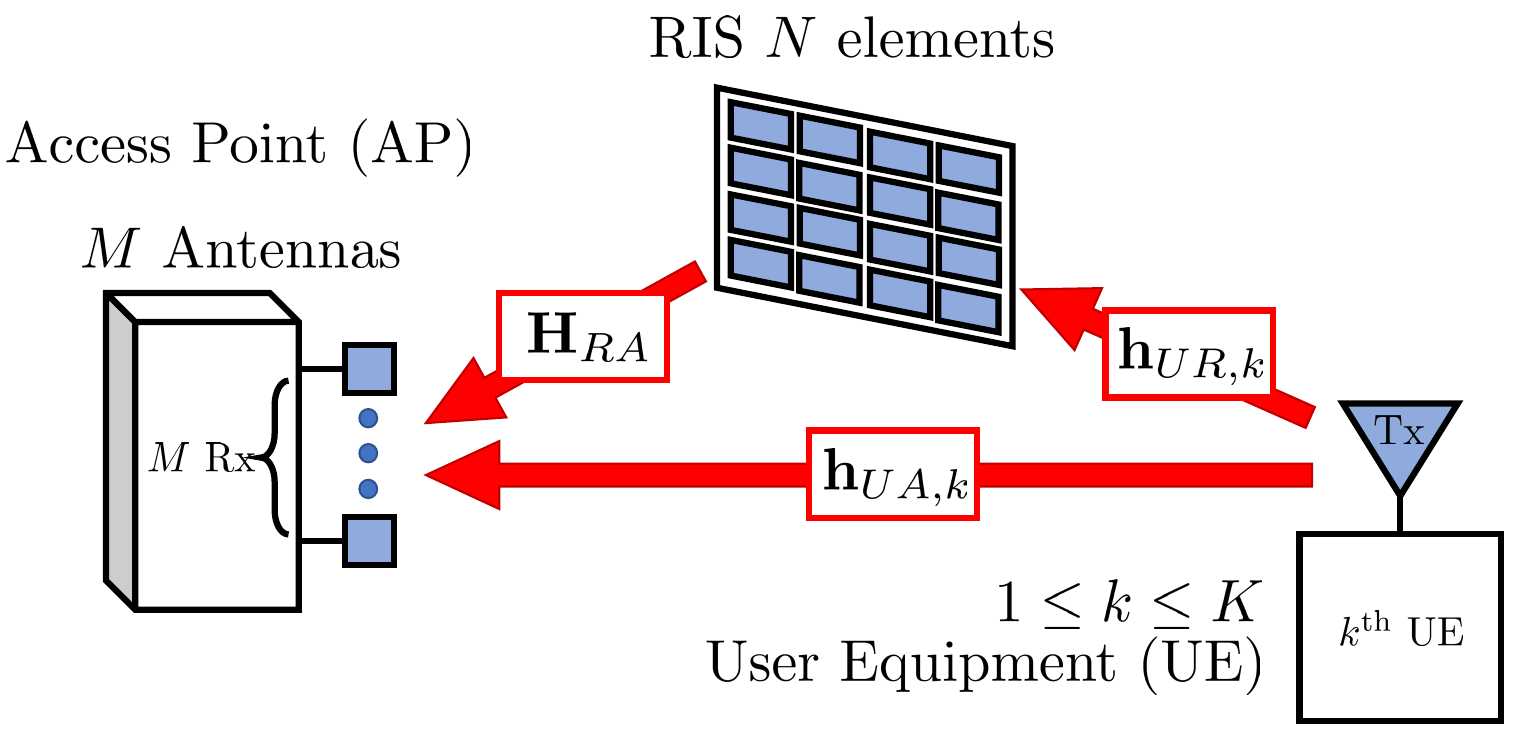}
    \caption{Uplink MIMO RIS communication model.}
    \label{fig:system-model}
\end{figure}

We consider an uplink narrowband MIMO RIS-assisted communication system comprising of an AP with $M$ antennas, $K$ single-antenna UEs, and an RIS with $N$ elements as shown in Fig.~\ref{fig:system-model}. 
The channels between the (A)ccess Point, (U)ser Equipment, and (R)IS are defined in Fig.~\ref{fig:system-model} and Table~\ref{tab:channel}.

In a time-slotted transmission scheme, the RIS phase shifts and all the transmitted pilot symbols from the UEs are functions of time $t\in\{1, \ldots, T\}$, where $T$ is the full training duration.
We assume all channels are block fading where they remain constant over $T$ time slots.
Let $\textbf{x}[t] \in \mathbb{C}^{K \times 1}$ be the vector containing the transmitted symbols of the $K$ UEs with $\mathbb{E}[\textbf{x}[t]\textbf{x}^H[t]] = P \textbf{I}_K$ where $P$ is the transmit power at each UE.
We assume that all $K$ UEs use the same transmit power.
The received signal at the AP $\textbf{y}[t]$ is:
\begin{align}
\label{eq:y[t]}
    \textbf{y}[t] = (\textbf{H}_{UA} + \textbf{H}_{RA} \diag(\boldsymbol{\phi}[t]) \textbf{H}_{UR}) \textbf{x}[t] + \textbf{n}[t]
\end{align}
where $\textbf{H}_{UA} = [\textbf{h}_{UA,1}, \ldots, \textbf{h}_{UA,K}]$, $\textbf{H}_{UR} =  [\textbf{h}_{UR,1}, \dots, \textbf{h}_{UR,K}]$, and $\textbf{n}[t] \sim \mathcal{CN}(\textbf{0}_{M \times 1}, \sigma^2 \textbf{I}_M)$ is the additive white Gaussian noise (AWGN) at the AP.
Finally, $\boldsymbol{\phi}[t] = [e^{j\theta_{1}[t]}, \ldots,  e^{j\theta_{N}[t]}]^T$  with $\theta_{n}[t] \in [-\pi, \pi)$ is the phase of the $n$th RIS element, $1 \le n \le N$.

The problem we consider in this work is to estimate the CSI at the AP comprising the channel matrices shown in Table~\ref{tab:channel} from $\textbf{y}[t]$ having knowledge of $\textbf{x}[t]$ and control of $\boldsymbol{\phi}[t]$.

\begin{table}[htbp]
\caption{Channel matrices estimated at the AP}
\begin{center}
\begin{tabular}{@{}lll@{}}\toprule
Description & Symbol & Size  \\ \midrule
Direct-path (UE to AP) & $\textbf{H}_{UA}$ & $M \times K$ \\
Cascaded channel (RIS to AP) & $\textbf{H}_{RA}$ & $M \times N$  \\
Cascaded channel (UE to RIS) & $\textbf{H}_{UR}$ & $N \times K$ \\ \bottomrule
\end{tabular}
\label{tab:channel}
\end{center}
\end{table}

\subsection{Block Transmission and Tensor Signal Modelling} 
\label{sec:block_tensor}
We now remodel the received signal (\ref{eq:y[t]}) into a third-order tensor using CP decomposition \cite{Kolda2009}.
We consider the total training duration of $T$ time slots to be partitioned into $B$ blocks, each containing $L$ time slots.
During each block $b \in \{1, \ldots, B\}$, the RIS phase values are constant and set to $\boldsymbol{\phi}[t] = \boldsymbol{\phi}[b]$, $t \in \{(b-1)L+1, \ldots, bL\}$.
From one block to the next, the RIS phase values are updated such that $\boldsymbol{\phi}^T[b] = [\boldsymbol{\Psi}]_{b,:}$, with $\boldsymbol{\Psi} \in \mathbb{C}^{B \times N}$.
Within each block, the $k$-th UE transmits $L$ pilots given by the $k$-th column vector of $\textbf{X} \in \mathbb{C}^{K \times L}$.
The matrix $\textbf{Y}[b] \in \mathbb{C}^{M \times L}$ collecting the received signals during the $b$-th block transmission is:
\begin{align}
    \textbf{Y}[b] = & (\textbf{H}_{UA} + \textbf{H}_{RA} \textbf{D}_{b}(\boldsymbol{\Psi}) \textbf{H}_{UR}) \textbf{X} + \textbf{N}[b] \nonumber \\
    = & \textbf{H}_{UA} \textbf{D}_{b}(\textbf{1}_{B \times K}) \textbf{X} + \textbf{H}_{RA} \textbf{D}_{b}(\boldsymbol{\Psi}) \textbf{H}_{UR} \textbf{X} + \textbf{N}[b] \label{eq:Y[b]}
\end{align}

We can reconstruct the received signal into a third-order tensor $\mathcal{Y} \in \mathbb{C}^{M \times L \times B}$ by using $\textbf{Y}[b]$ as the $b$-th frontal face matrix of $\mathcal{Y}$.
Using short form notation for CP decompositions with the factor matrices \cite{Kolda2009}, we can write the received signal tensor using two tensor CP decompositions:
\begin{flalign}
    \label{eq:Ytensor}
    \mathcal{Y} = & [\![\textbf{H}_{UA}, \textbf{X}^T, \textbf{1}_{B \times K} ]\!] + [\![\textbf{H}_{RA}, \textbf{Z}^T, \boldsymbol{\Psi} ]\!] + \mathcal{N}
\end{flalign}
where $\textbf{Z} = \textbf{H}_{UR}\textbf{X} \in \mathbb{C}^{N \times L}$.
The tensor signal is composed of two CP decompositions: $[\![\textbf{H}_{UA}, \textbf{X}^T, \textbf{1}_{B \times K} ]\!]$ which corresponds to the direct-path channels and $[\![\textbf{H}_{RA}, \textbf{Z}^T, \boldsymbol{\Psi} ]\!]$ for the path that goes through the RIS.
These factor matrices are summarized in Table~\ref{tab:factormatrices}.

\begin{table}[htbp]
\caption{Factor matrices of the CP decompositions in (\ref{eq:Ytensor})}
\begin{center}
\begin{tabular}{@{}lll@{}}\toprule
CP decomposition term & Factor matrix & Size  \\ \midrule
\multirow{3}{*}{Non-RIS-assisted link} & $\textbf{H}_{UA}$ & $M \times K$  \\
                     & $\textbf{X}^T$ & $L \times K$ \\ 
                     & $\textbf{1}_{B \times K}$ & $B \times K$ \\ \midrule
\multirow{3}{*}{RIS-assisted link} & $\textbf{H}_{RA}$ & $M \times N$ \\ 
                     & $\textbf{Z}^T$ & $L \times N$ \\ 
                     & $\boldsymbol{\Psi}$ & $B \times N$ \\ \bottomrule
\end{tabular}
\label{tab:factormatrices}
\end{center}
\end{table}

\section{Channel Estimation}
\label{sec:channel_estimation}
To estimate the CSI at the AP we propose two methods.
The first method uses two transmission channel estimation stages.
The second method improves upon the first method to jointly estimate all channels using a single pilot transmission stage.

\subsection{Two Stage RIS OFF-ON  Channel Estimation} 
We propose to estimate the CSI by turning the RIS OFF and estimate the direct path in the first stage, then turn the RIS ON and subtract the estimated direct path from the receive signal to estimate the cascaded channels in the second stage.

\subsubsection{Stage 1 -- Estimation of the Direct Path Channels}
In the first stage, we turn the RIS OFF for $L'$ time slots and we transmit pilots $\bar{\textbf{X}} \in \mathbb{C}^{K \times L'}$ from the UEs.
The received signal $\textbf{V} \in \mathbb{C}^{M \times L'}$ at the AP is:
\begin{flalign}
    \textbf{V} = \textbf{H}_{UA} \bar{\textbf{X}} + \textbf{N} 
\end{flalign}
and the LS estimate of the direct path channel $\hat{\textbf{H}}_{UA}$ is:
\begin{flalign}
    \label{eq:Gstage1}
    \hat{\textbf{H}}_{UA} =& \arg\min_{\textbf{H}_{UA}} \|\textbf{V} - \textbf{H}_{UA}\bar{\textbf{X}}\|^2_F = \textbf{V} \bar{\textbf{X}}^\dagger_{\text{right}}
\end{flalign}

\subsubsection{Stage 2 -- Estimation of the Cascaded Channel with Alternating Least Squares}
In the second stage, the RIS is turned ON and we transmit from the UEs $B$ block pilot sequences, where each block contains $L$ pilots.
We can remove the direct path terms from the $b$-th block transmission in (\ref{eq:Y[b]}) using the estimated channels in Stage 1 to get the received signal going through the RIS:
\begin{flalign}
    \label{eq:Q[b]}
    \textbf{Q}[b] =& \textbf{Y}[b] - \hat{\textbf{H}}_{UA}\textbf{X} = \textbf{H}_{RA} \textbf{D}_{b}(\boldsymbol{\Psi}) \textbf{Z} + \bar{\textbf{N}}[b]
\end{flalign}
where $\bar{\textbf{N}}[b] = \textbf{N}[b] + (\textbf{H}_{UA} - \hat{\textbf{H}}_{UA})\textbf{X}$ is the effective noise due to the error of estimating $\textbf{H}_{UA}$ in Stage 1.
The remaining portion of the received signal $\textbf{Q}[b]$ corresponds to the matrix frontal slices of a third-order tensor with a CP decomposition of the RIS channels, pilots, and the RIS phase shift matrix as $\mathcal{Q} = [\![\textbf{H}_{RA}, \textbf{Z}^T, \boldsymbol{\Psi}]\!]\in \mathbb{C}^{M \times L \times B}$.
We can obtain the unfoldings for $\mathcal{Q}$ based on \cite{Kolda2009} as:
\begin{flalign}
    \textbf{Q}_{1} = \textbf{H}_{RA}(\boldsymbol{\Psi} \diamond \textbf{Z}^T)^T + \bar{\textbf{N}}_{1} \label{eq:Q1} \\
    \textbf{Q}_{2} = \textbf{Z}^T (\boldsymbol{\Psi} \diamond \textbf{H}_{RA})^T + \bar{\textbf{N}}_{2} \label{eq:Q2} 
\end{flalign}

With the first two unfoldings above, we can derive an ALS algorithm based on \cite{Kolda2009} to obtain individual estimates of the RIS-assisted channel matrices $\hat{\textbf{H}}_{UR}$ and $\hat{\textbf{H}}_{RA}$ by alternating between the two following minimization problems:
\begin{flalign}
    \hat{\textbf{H}}_{RA} = \arg\min_{\textbf{H}_{RA}} \|\textbf{Q}_1 - \textbf{H}_{RA}(\boldsymbol{\Psi} \diamond \textbf{Z}^T)^T\|^2_F \\
    \hat{\textbf{Z}} = \arg\min_{\textbf{Z}} \|\textbf{Q}_2 - \textbf{Z}^T (\boldsymbol{\Psi} \diamond \textbf{H}_{RA})^T\|^2_F
\end{flalign}
whose solutions are:
\begin{flalign}
    \hat{\textbf{H}}_{RA} = & \textbf{Q}_1 ((\boldsymbol{\Psi} \diamond \textbf{Z}^T)^T)^\dagger_{\text{right}} \label{eq:ALS-HRA} \\
    \hat{\textbf{Z}} = & (\boldsymbol{\Psi} \diamond \textbf{H}_{RA})^\dagger_{\text{left}} \textbf{Q}_2^T \label{eq:ALS-Z}
\end{flalign}

The algorithm terminates when the normalized squared Frobenius norm of the difference between two consecutive iterations of $\hat{\textbf{Z}}$ and $\hat{\textbf{H}}_{RA}$ are less than a threshold $\delta$ or when the maximum number of iterations is exceeded (line 7 in Algorithm~\ref{alg:ALS}).
After the algorithm converges, we obtain our estimate of $\hat{\textbf{H}}_{UR} = \hat{\textbf{Z}}\textbf{X}^\dagger_\text{right}$.
The iterative ALS algorithm is described in Algorithm~\ref{alg:ALS}. 

\begin{algorithm}
\small
\label{alg:ALS}
% \setstretch{1.5}
\flushleft
\caption{Alternating Least Squares (ALS)} \label{alg:ALS}
\begin{algorithmic}[1]
\State \textbf{input:} $\textbf{X}$, $\mathcal{Q}$
\State \textbf{initialize:} randomly generate $\hat{\textbf{H}}_{UR}$, $\hat{\textbf{H}}_{RA}$, set algorithmic iteration as $i=1$.
$$\hat{\textbf{Z}}_{(i=0)} \gets \hat{\textbf{H}}_{UR}\textbf{X}$$
$$\hat{\textbf{H}}_{RA(i=0)} \gets \hat{\textbf{H}}_{RA}$$
$$\textbf{Q}_1 \gets \begin{bmatrix}\mathcal{Q}_{:,:,1}& \ldots& \mathcal{Q}_{:,:,B}\end{bmatrix}$$
$$\textbf{Q}_2 \gets \begin{bmatrix}\mathcal{Q}_{:,:,1}^T& \ldots& \mathcal{Q}_{:,:,B}^T\end{bmatrix}$$
\Repeat
\State $\hat{\textbf{H}}_{RA(i)} \gets \textbf{Q}_{1} ((\boldsymbol{\Psi} \diamond \hat{\textbf{Z}}^T_{(i-1)})^T)^\dagger_{\text{right}}$
\State $\hat{\textbf{Z}}_{(i)} \gets (\boldsymbol{\Psi} \diamond \hat{\textbf{H}}_{RA(i)} )^\dagger_{\text{left}} \textbf{Q}_{2}^T$
\State $i \gets i + 1$
\Until{$\|\hat{\textbf{Z}}_{(i)} - \hat{\textbf{Z}}_{(i-1)}\|^2_F \|\hat{\textbf{Z}}_{(i)}\|^{-2}_F \le \delta$ and $\|\hat{\textbf{H}}_{RA(i)} - \hat{\textbf{H}}_{RA(i-1)}\|^2_F \|\hat{\textbf{H}}_{RA(i)}\|^{-2}_F \le \delta$ or $i > I_{\text{max}}$}
\State $\hat{\textbf{H}}_{UR} \gets \hat{\textbf{Z}}_{(i)} \textbf{X}^\dagger_{\text{right}}$
\State $\hat{\textbf{H}}_{RA} \gets \hat{\textbf{H}}_{RA(i)}$
\State \Return $\hat{\textbf{H}}_{UR}$, $\hat{\textbf{H}}_{RA}$
\end{algorithmic}
\end{algorithm}

\subsection{Enhanced Alternating Least Squares Channel Estimation}
For our second proposed method, we enhance the traditional ALS algorithm \cite{Kolda2009} to estimate the factor matrices of our tensor signal model (\ref{eq:Ytensor}) which is composed of two CP decompositions instead of one.
We can obtain the unfoldings for $\mathcal{Y}$ as:
\begin{flalign}
    \textbf{Y}_{1} = \textbf{H}_{UA} (\textbf{1} \diamond \textbf{X}^T)^T + \textbf{H}_{RA}(\boldsymbol{\Psi} \diamond \textbf{Z}^T)^T + \textbf{N}_{1} \label{eq:Y1} \\
    \textbf{Y}_{2} = \textbf{X}^T (\textbf{1} \diamond \textbf{H}_{UA})^T + \textbf{Z}^T (\boldsymbol{\Psi} \diamond \textbf{H}_{RA})^T + \textbf{N}_{2} \label{eq:Y2}
\end{flalign}
where $\textbf{1} = \textbf{1}_{B \times K}$ in (\ref{eq:Y1}) and (\ref{eq:Y2}).
We can see that the noise terms of these unfoldings are only composed of AWGN compared to the two-stage method in (\ref{eq:Q1}) and (\ref{eq:Q2}).

From (\ref{eq:Y1}) and (\ref{eq:Y2}) we can estimate the CSI by alternating between the two following optimization problems:
\begin{flalign}
    \arg\min_{\textbf{H}_{UA},\textbf{H}_{RA}} \left\| \textbf{Y}_{1} - \begin{bmatrix} \textbf{H}_{UA} & \textbf{H}_{RA} \end{bmatrix} \begin{bmatrix} (\textbf{1} \diamond \textbf{X}^T)^T \\ (\boldsymbol{\Psi} \diamond \textbf{Z}^T)^T \end{bmatrix} \right\|^2_F \label{eq:minHUAHRA} \\
    \arg\min_{\textbf{Z}} \left\| \textbf{Y}_{2} - \textbf{X}^T (\textbf{1} \diamond \textbf{H}_{UA})^T - \textbf{Z}^T (\boldsymbol{\Psi} \diamond \textbf{H}_{RA})^T \right\|^2_F \label{eq:minZ}
\end{flalign}
whose solutions are:
\begin{align}
    \begin{bmatrix} \hat{\textbf{H}}_{UA} & \hat{\textbf{H}}_{RA} \end{bmatrix} =
\textbf{Y}_{1} \begin{bmatrix} (\textbf{1} \diamond \textbf{X}^T)^T \\ (\boldsymbol{\Psi} \diamond \textbf{Z}^T)^T \end{bmatrix}^\dagger_{\text{right}} \label{eq:E-ALS-HUAHRA} \\
    \hat{\textbf{Z}} = (\boldsymbol{\Psi} \diamond \textbf{H}_{RA} )^\dagger_{\text{left}} (\textbf{Y}_{2}^T - (\textbf{1} \diamond \textbf{H}_{UA} ) \textbf{X}) \label{eq:E-ALS-Z}
\end{align}
where $\textbf{1} = \textbf{1}_{B \times K}$ in (\ref{eq:minHUAHRA}), (\ref{eq:minZ}), (\ref{eq:E-ALS-HUAHRA}), and (\ref{eq:E-ALS-Z}).
Similarly to Algorithm~\ref{alg:ALS}, we obtain our estimate of $\hat{\textbf{H}}_{UR} = \hat{\textbf{Z}}\textbf{X}^\dagger_\text{right}$ after the algorithm converges.
The proposed iterative E-ALS algorithm is described in Algorithm~\ref{alg:E-ALS}.

\begin{algorithm}
\small
\label{alg:E-ALS}
% \setstretch{1.5}
\flushleft
\caption{Enhanced Alternating Least Squares (E-ALS)} \label{alg:E-ALS}
\begin{algorithmic}[1]
\State \textbf{input:} $\textbf{X}$, $\mathcal{Y}$
\State \textbf{initialize:} randomly generate $\hat{\textbf{H}}_{UA}$, $\hat{\textbf{H}}_{UR}$, $\hat{\textbf{H}}_{RA}$, set algorithmic iteration as $i=1$.
$$\hat{\textbf{Z}}_{(i=0)} \gets \hat{\textbf{H}}_{UR}\textbf{X}$$
$$\hat{\textbf{H}}_{UA(i=0)} \gets \hat{\textbf{H}}_{UA}$$
$$\hat{\textbf{H}}_{RA(i=0)} \gets \hat{\textbf{H}}_{RA}$$
$$\textbf{Y}_1 \gets \begin{bmatrix}\mathcal{Y}_{:,:,1}& \ldots& \mathcal{Y}_{:,:,B}\end{bmatrix}$$
$$\textbf{Y}_2 \gets \begin{bmatrix}\mathcal{Y}_{:,:,1}^T& \ldots& \mathcal{Y}_{:,:,B}^T\end{bmatrix}$$
\Repeat
\State $\begin{bmatrix}\hat{\textbf{H}}_{UA(i)} & \hat{\textbf{H}}_{RA(i)} \end{bmatrix} \gets \textbf{Y}_{1} \begin{bmatrix} (\textbf{1} \diamond \textbf{X}^T)^T \\ (\boldsymbol{\Psi} \diamond \hat{\textbf{Z}}^T_{(i-1)})^T \end{bmatrix}^\dagger_{\text{right}}$
\State $\hat{\textbf{Z}}_{(i)} \gets (\boldsymbol{\Psi} \diamond \hat{\textbf{H}}_{RA(i)} )^\dagger_{\text{left}} (\textbf{Y}_{2}^T - (\textbf{1} \diamond \hat{\textbf{H}}_{UA(i)} ) \textbf{X})$
\State $i \gets i + 1$
\Until{$\|\hat{\textbf{H}}_{UA(i)} - \hat{\textbf{H}}_{UA(i-1)}\|^2_F \|\hat{\textbf{G}}_{(i)}\|^{-2}_F \le \delta$ and $\|\hat{\textbf{Z}}_{(i)} - \hat{\textbf{Z}}_{(i-1)}\|^2_F \|\hat{\textbf{Z}}_{(i)}\|^{-2}_F \le \delta$ and $\|\hat{\textbf{H}}_{RA(i)} - \hat{\textbf{H}}_{RA(i-1)}\|^2_F \|\hat{\textbf{H}}_{RA(i)}\|^{-2}_F \le \delta$ or $i > I_{\text{max}}$}
\State $\begin{bmatrix}\hat{\textbf{H}}_{UA} & \hat{\textbf{H}}_{RA} \end{bmatrix} \gets \begin{bmatrix} \hat{\textbf{H}}_{UA(i)} & \hat{\textbf{H}}_{RA(i)} \end{bmatrix}$
\State $\hat{\textbf{H}}_{UR} \gets \hat{\textbf{Z}}_{(i)} \textbf{X}^\dagger_{\text{right}}$
\State \Return $\hat{\textbf{H}}_{UA}$, $\hat{\textbf{H}}_{UR}$, $\hat{\textbf{H}}_{RA}$
\end{algorithmic}
\end{algorithm}

\subsection{Computational Complexity}
We summarize the computational complexity of the proposed channel estimation methods in Table~\ref{tab:compcomplex}.
For the two-stage RIS OFF-ON method, in the first stage, the complexity of calculating $\hat{\textbf{H}}_{UA}$ comes from matrix multiplication where the pseudo inverse term $\bar{\textbf{X}}^{\dagger}_{\text{right}}$  can be determined before run-time. 
In the second stage the complexity of calculating $\hat{\textbf{H}}_{RA(i)}$ (\ref{eq:ALS-HRA}) and $\hat{\textbf{Z}}_{(i)}$ (\ref{eq:ALS-Z}) occurs over every iteration in Algorithm~\ref{alg:ALS} (line 4 and 5).
For the E-ALS method, all channels are estimated based on $\hat{\textbf{H}}_{(i)} = \begin{bmatrix}\hat{\textbf{H}}_{UA(i)} & \hat{\textbf{H}}_{RA(i)} \end{bmatrix}$ (\ref{eq:E-ALS-HUAHRA}) and $\hat{\textbf{Z}}_{(i)}$ (\ref{eq:E-ALS-Z}) throughout each iteration in Algorithm~\ref{alg:E-ALS} (line 4 and 5).

\begin{table}[htbp]
\scriptsize
\caption{Computational Complexity of our proposed channel estimation methods}
\begin{center}
\begin{tabular}{@{}ll@{}}\toprule \midrule
Two Stage & Computational Complexity $\mathcal{O}(\cdot)$ \\ \midrule
$\hat{\textbf{H}}_{UA}$ & $MKL'$ \\
$\hat{\textbf{H}}_{RA(i)}$ & $N^3 + N^2BL + NBL(BL + M + 1)$ \\
$\hat{\textbf{Z}}_{(i)}$ & $N^3 + N^2ML + NML(ML + B + 1)$ \\ \midrule \\ \midrule
E-ALS  & Computational Complexity $\mathcal{O}(\cdot)$ \\ \midrule
$\hat{\textbf{H}}_{(i)}$ & $(N+K)^3 + (N+K)^2BL + (N+K)(BL + M + 1)BL$ \\ 
$\hat{\textbf{Z}}_{(i)}$ & $N^3 + N^2BM + NBM(BM + L + 1) + BM(K + KL + L)$ \\ \midrule \bottomrule
\end{tabular}
\label{tab:compcomplex}
\end{center}
\end{table}

\section{Simulations}
\label{sec:simulations}
\subsection{Simulation Setup}
We simulate a system with $M=4$ transmit and receive antennas at the AP configured in a uniform linear array (ULA), $K=8$ single-antenna UEs, and $N=5 \times 5=25$ elements for a uniform rectangular array (URA) RIS. 
The signal-to-noise ratio (SNR) is defined as $P/\sigma^2$.
For Algorithms \ref{alg:ALS} and \ref{alg:E-ALS} we set the maximum number of iterations $I_{max} = 20$ and the convergence threshold $\delta = 10^{-8}$.

All channels are modeled with large-scale fading parameters and distances adopted from \cite{Fernandes2023}. 
The pathloss expressed in dB is $\rho = \rho_0({d}/{d_0})^{-\alpha}$ where $\rho_0 = -20 \text{ dB}$, $d$ is the distance, $d_0$ is the reference distance of 1m, and $\alpha$ is the path loss exponent.
Specifically, for the AP-RIS link we use $d_{AR} = 20\text{m}$ with $\alpha_{AR} = 2.1$, for the UE-RIS link $d_{UR} = 20\text{m}$ with $\alpha_{UR} = 4.2$, and for the UE-AP link $d_{UA} = 30\text{m}$ with $\alpha_{UA} = 2.2$. 

All channels are modelled geometrically using steering vectors with angles of arrival (AoA) and angles of departure (AoD) \cite{Zheng2014a,Chen2021} with $R=2$ paths, given by:
\begin{flalign}
    \textbf{H}_{RA} =& \sqrt{\rho_{RA}} \sum_{i=1}^{R} \alpha_{RA_{i}} \textbf{a}_{ULA}(\phi_{RA_{i}}) \textbf{a}_{URA}^H(\theta_{RA_{i}}, \psi_{RA_{i}}) \\
    % \textbf{H}_{UR} =& \begin{bmatrix}\textbf{h}_{UR,1} & \ldots & \textbf{h}_{UR,K}\end{bmatrix} \\
    \textbf{h}_{UR,k} =& \sqrt{\rho_{UR}} \sum_{i=1}^{R} \alpha_{UR_{i,k}} \textbf{a}_{URA}(\theta_{UR_{i,k}},\psi_{UR_{i,k}}) \\
    % \textbf{H}_{UA} =& \begin{bmatrix}\textbf{h}_{UA,1} & \ldots & \textbf{h}_{UA,K}\end{bmatrix} \\
    \textbf{h}_{UA,k} =& \sqrt{\rho_{UA}} \sum_{i=1}^{R} \alpha_{UA_{i,k}} \textbf{a}_{ULA}(\theta_{UA_{i,k}})
\end{flalign}
where $\alpha_x \sim \mathcal{CN}(0,1)$, $x \in\{RA_{i}, UR_{i,k}, UA_{i,k}\}$, describe the small scale fading.
The general equations of steering vectors for a ULA $\textbf{a}_{ULA}(\theta)$ and a URA $\textbf{a}_{URA}(\psi,\phi)$ with $M$ antennas are
$
\textbf{a}_{ULA}(\phi) = \begin{bmatrix}1 & e^{j\frac{2\pi}{\lambda} l \sin{\theta}} & \ldots & e^{j\frac{2\pi}{\lambda} l (M-1) \sin{\theta}} \end{bmatrix}
$ and 
$
    \textbf{a}_{URA}(\theta,\psi) = \textbf{a}_y(\psi,\phi) \otimes \textbf{a}_x(\psi,\phi)
$
where all AoAs and AoDs are uniformly distributed based on \cite{Chen2021}, where at the AP $\phi \in [0, \frac{\pi}{2})$, while at the RIS $\theta \in [0, \frac{\pi}{2})$ (elevation angle) and $\psi \in [0, \pi)$ (azimuth angle). 
The distance between adjacent antennas is assumed to be $l=\frac{\lambda}{2}$, the vertical and horizontal $\textbf{a}_x(\theta,\psi)$ directions for the URA respectively are $\textbf{a}_y(\theta,\psi) = \begin{bmatrix}1 & e^{j\frac{2\pi}{\lambda} l \sin{\theta}\sin{\psi}} & \ldots & e^{j\frac{2\pi}{\lambda} l (M-1) \sin{\theta}\sin{\psi}} \end{bmatrix}^T$ and $\textbf{a}_x(\theta,\psi) = \begin{bmatrix}1 & e^{j\frac{2\pi}{\lambda} l \sin{\theta}\cos{\psi}} & \ldots & e^{j\frac{2\pi}{\lambda} l (M-1) \sin{\theta}\cos{\psi}} \end{bmatrix}^T$.

The performance of the channel estimators is measured by the normalized mean square error (NMSE) defined as
$\text{NMSE}(\hat{\textbf{H}}) = \frac{\|\textbf{H} - \hat{\textbf{H}}\|^2_F}{\|\textbf{H}\|^2_F}
$.
To account for scaling ambiguity of the decoupled cascaded channels since $\hat{\textbf{H}}_{RA}\hat{\textbf{H}}_{UR} = \hat{\textbf{H}}_{RA}\boldsymbol{\Delta}_{RA} \boldsymbol{\Delta}_{UR}\hat{\textbf{H}}_{UR}$, where $\boldsymbol{\Delta}_{RA}\boldsymbol{\Delta}_{UR} = \textbf{I}_N$, we normalize the first column of the channel matrices when calculating the NMSE \cite{Wei2021}.
These scaling ambiguities are irrelevant for evaluation as they compensate for each other when constructing the estimated cascaded channels $\hat{\textbf{H}}_{RA}\hat{\textbf{H}}_{UR}$.

When comparing the channel estimation methods, we ensure that the total training duration is the same across all methods during each block transmission.
When the RIS is turned OFF for the first stage of the two-stage RIS OFF-ON method, we make $L' = L$ (the length of one block transmission).
Then, for the second stage $\boldsymbol{\Psi} = \textbf{F}_{N}$ (i.e., $B=N$), where $\textbf{F}_{N}$ is an $N\times N$-sized DFT matrix for Algorithm \ref{alg:ALS}. 
For the E-ALS and LS methods, a choice for the RIS phase shifts is to adopt the DFT RIS phase shift method from \cite{Jensen2020}, by making $\begin{bmatrix} \textbf{1}_{N+1 \times 1} & \boldsymbol{\Psi} \end{bmatrix} = \textbf{F}_{N+1}$ (i.e., $B = N+1$) for Algorithm \ref{alg:E-ALS}.
Since both methods take the pseudoinverse of $\textbf{X}$ to obtain $\hat{\textbf{H}}_{UR}$ from $\hat{\textbf{Z}}$, we transmit orthogonal pilots such that $\textbf{X}$ is constructed out of a DFT matrix, transmitting $L = K$ pilots per block.
The total training duration for the two stage RIS OFF-ON method of $T = L' + BL = (N+1)L$ will be the same as the E-ALS method of $T = BL = (N+1)L$ at $T = 208$.

\subsection{Numerical Results}
\begin{figure}[tb]
    \centering
    \includegraphics[width=0.8\columnwidth]{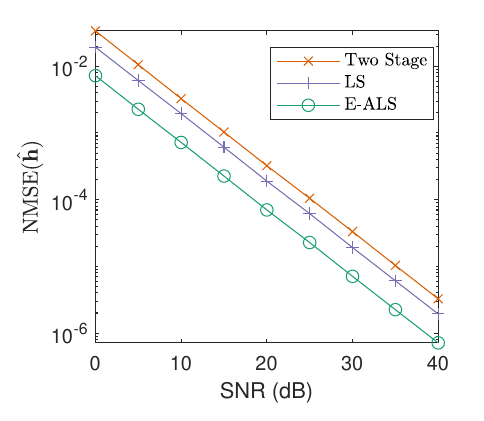}
    \caption{NMSE vs SNR of the enhanced alternating least squares, two stage RIS OFF-ON, and least squares channel estimation methods.}
\label{fig:NMSEvsSNR_allCEmethods}
\end{figure}
\begin{figure}[tb]
    \centering
    \includegraphics[width=\columnwidth]{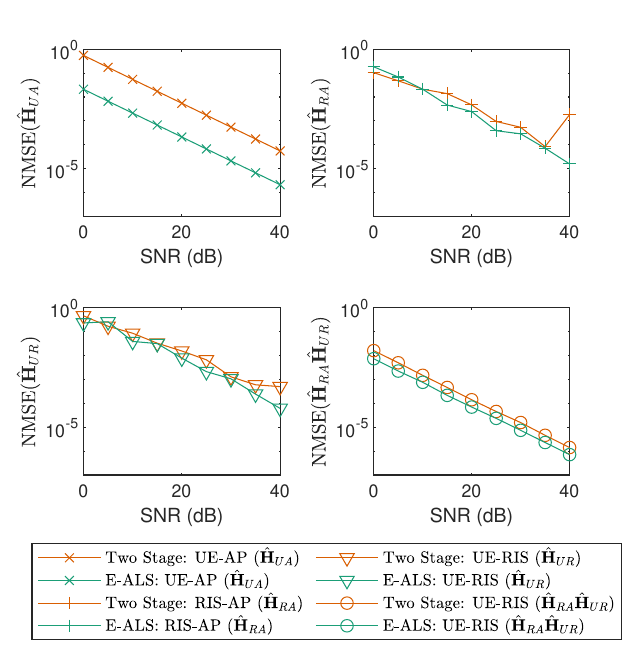}
    \caption{NMSEs vs SNR of each of the individual channel matrices and the reconstructed cascaded channels.}
    \label{fig:NMSEvsSNR_IndividualChannels}
\end{figure}
\begin{figure}[tb]
    \centering
    \includegraphics[width=0.8\columnwidth]{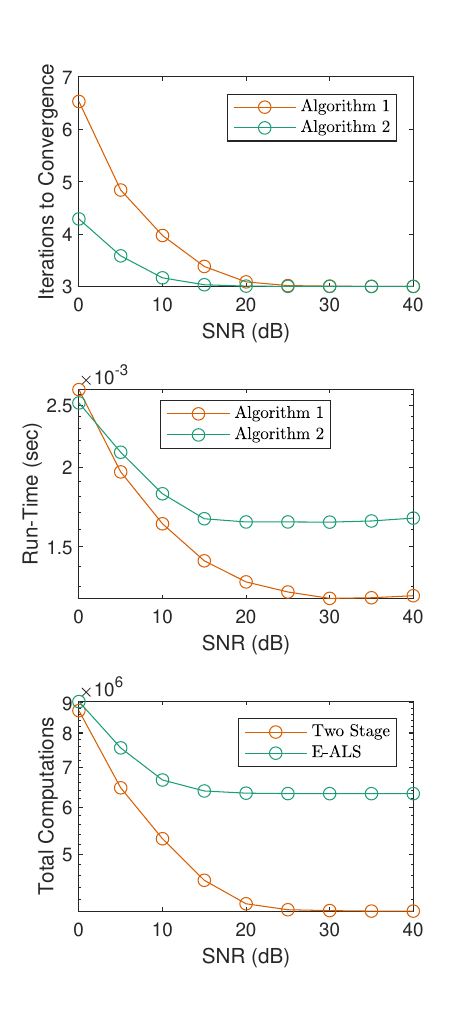}
    \caption{Average number of iterations to convergence, average run-time, and total number of computations vs SNR for Algorithms~\ref{alg:ALS} and~\ref{alg:E-ALS} in each channel estimation method.}
    \label{fig:IterationsvsSNR_EALSvsALS}
\end{figure}

All results presented are averages over 10000 Monte-Carlo simulations. Fig.~ \ref{fig:NMSEvsSNR_allCEmethods} shows a comparison between each channel estimation method where all CSI is estimated as a parameter vector $\hat{\bold{h}} = [\text{vec}(\hat{\textbf{H}}_{UA})^T, \text{vec}(\hat{\textbf{H}}_{UR}^T \diamond \hat{\textbf{H}}_{RA})^T]^T$ for comparison with the LS method \cite{Swindlehurst2022}.
From most accurate to least is the E-ALS, the LS, then the two stage RIS OFF-ON.
These results show that the E-ALS method is able to exploit the structure of our tensor signal model, whereas using two separate stages gives a less accurate estimate than the traditional LS solution.

Figure~\ref{fig:NMSEvsSNR_IndividualChannels} shows the NMSE vs SNR of the individual channel matrices estimated by the two-stage RIS OFF-ON and E-ALS channel estimation methods.
We can see a gap between the NMSE of the two-stage RIS OFF-ON and E-ALS methods for all the individual channels because the E-ALS method uses all pilots efficiently to estimate all channels throughout the full training duration.

Figure~\ref{fig:IterationsvsSNR_EALSvsALS} shows the average number of iterations and average run-time for Algorithm \ref{alg:ALS} (ALS) and Algorithm \ref{alg:E-ALS} (E-ALS) to converge along with the total number of computations to estimate all channels for each method (calculated based on the iterations and the computational complexity in Table \ref{tab:compcomplex}).
The E-ALS algorithm converges more quickly with fewer iterations but it still uses more computations with a longer average run-time.
The E-ALS method has a higher run-time and computational complexity due to the dominating factor of $\mathcal{O}((N+K)^3)$ when jointly computing both $\hat{\textbf{H}}_{UA}$ and $\hat{\textbf{H}}_{RA}$ on line 4 of Algorithm~\ref{alg:E-ALS} compared to $\mathcal{O}(N^3)$ when computing only $\hat{\textbf{H}}_{RA}$ on line 4 of Algorithm~\ref{alg:ALS} for the two stage method.

\section{Summary and Conclusion}
\label{sec:conclusion}
We proposed two channel estimation methods (two stage RIS OFF-ON and E-ALS) for narrowband RIS-assisted MIMO communication systems that exploited tensor signal modelling techniques to estimate the direct-path, and RIS-assisted channels.
We derived a tensor signal model with two CP decomposition terms from the received signal for our proposed channel estimation methods and compared them with the traditional LS method.
Through simulations we showed that both of our proposed channel estimation methods were able to obtain decoupled estimates of the cascaded RIS channels.
Simulations showed that the E-ALS method efficiently used the total training duration and structure of our tensor signal model to achieve a more accurate estimate than the two stage RIS OFF-ON and traditional LS methods at the cost of a higher computational complexity.

% {\appendices % begin Appendices
% \section{}
% } % end Appendices

\end{document}